\documentclass[11pt]{article}

\usepackage{graphicx}
\usepackage{dcolumn}   
\usepackage{amsmath} 
\usepackage{epsfig}
\usepackage{hhline}
\RequirePackage{xspace}
\usepackage{bm}
\begin{document}
\bibliographystyle{unsrt}

\pagestyle{empty}               
	
\rightline{\vbox{
	\halign{&#\hfil\cr
	&RHIC Spin Note\cr}}}	

\rightline{\vbox{
	\halign{&#\hfil\cr
	&May 2003\cr}}}
\vskip 1in
\begin{center}
{\Large\bf
{Proton-carbon CNI polarimetry and the spin-dependence of the Pomeron}}

\vskip .5in
\normalsize
T.L.\ Trueman \footnote{This manuscript has been authored
under contract number DE-AC02-76CH00016 with the U.S. Department
of Energy.  Accordingly, the
U.S. Government retains a non-exclusive, royalty-free license to
publish or reproduce the published form of this contribution, or
allow others to do so, for U.S. Government purposes.}\\
{\sl Physics Department, Brookhaven National 
Laboratory, Upton, NY 11973}							
\end{center}
\vskip 0.5in
\begin{abstract} 
  Recent polarized proton experiments at Brookhaven National Laboratory are used as a basis for a model of the energy dependence of the analyzing power of proton-carbon elastic scattering.  In addition to their practical value for polarimetry,  the results of this analysis give  constraints on the size of the Pomeron spin-flip coupling as well as information on the $f_2$ and $\omega$ spin-flip couplings.\end{abstract}
\vfill \eject \pagestyle{plain}
\setcounter{page}{1}
\section{} 
During the past three years experiments have been carried out at Brookhaven
National Laboratory, at the AGS with proton beam momentum of 21.7 GeV/c and
at RHIC with beam momenta of~24~GeV/c and 100 GeV/c, to measure the
single-spin asymmetry for elastic scattering of transversely polarized
proton beams off (fixed) carbon targets. The principal goal of these
experiments is to obtain a calibrated polarimeter based on elastic $pC$
scattering in the CNI region, i.e. $|t| \le 0.05$~${\rm (GeV/c)}^2$. In outline, the
procedure to be used is to begin with an AGS beam of known polarization, and to
measure the analyzing power $A_N(t)$ near the RHIC injection energy. In the CNI
region the analyzing power is expected to be a few percent which, coupled with the
large rate, makes it well-suited to be a practical polarimeter. The ultimate goal 
is to have a sufficently precise calibration to
make the desired 5\% measurement of the polarization $P$.  (As we will see the current experimental errors are too large to reach this level of absolute accuracy, though the relative errors are quite small.. This should improve soon as further experiments are carried out,  but it may ultimately require the use of a polarized hydrogen gas jet target to reach the required level of 5\% absolute). 
This AGS experiment is referred to as E950. See \cite{E950} for a description of the
experiment and the results. The next step is to measure the asymmetry after transfer
of the beam into RHIC with a setup the same as or similar to the one in the AGS. The
energy at this point is a little higher than where the calibration is made, but the
analyzing power is not expected to change enough that significant additional
uncertainty is introduced into the measurement of $P$.

The difficult and as yet not solved problem has to do with the polarization at
higher energy in RHIC. It would be nice to believe that the polarization is not reduced
during the RHIC acceleration, but this is not really known, and it would be good to
have some theoretical guidance as to what sort of energy dependence the CNI
analyzing power is expected to have. The central goal of this paper is to present a
plausible model, based on Regge phenomenology, to calculate $A_N(t)$ at higher
energy assuming only that a calibrated measurement can be made at injection. The
model, as utilized here, has as spin-off some interesting theoretical results. In
particular, the question of the spin-dependence ot the Pomeron coupling to the
proton has been studied for many years with still no firm conclusion. (See
\cite{Buttimore} for an extensive discussion of this issue.) It is generally known
to be small, less than 10-15\%, but not better known than that. The analyzing power
is very sensitive to this coupling, as we will see, and so even low energy data,
like that from the AGS, can give us very good information about this coupling.

The
model we will adopt uses three Reggeons, the Pomeron
$P$ and two Regge poles,
$f_2$ and
$\omega$. Because carbon is an isosinglet none of the $I=1$ Regge poles like the
$\rho$ or
$a_2$ can enter, but because they are known to make small contributions to the elastic, non-flip amplitudes \cite{Berger} we can use the fits for unpolarized $pp$
and  $\bar{p} p$ elastic scattering that have been carried out over an extensive energy
range. \cite{Cudell,Block} (To carry over our results to $pp$ spin-flip will require
important corrections because $\rho$ and
$a_2$ spin-flip couplings are known to be strong
\cite{Berger, Kramer}. See Section 5, below.)

We will begin by discussing the determination of $A_N(t)$ from the data obtained in
E950. This analyzing power will then be applied to determine the polarization in
RHIC from the asymmetry measured there at 24 GeV/c. Next we will describe how the
3-Reggeon model is extended from elastic non-flip scattering to spin-flip scattering
and will examine what we can say about the model just based on the E950 results. We
will then examine the RHIC data taken at 100 GeV/c and will see that, without
knowing the polarization in RHIC at that energy,   all the parameters of our model can be
determined by measuring, in addition to the E950 results, the polarization-independent
``shape" of the asymmetry distribution in the CNI region. This being done, we can
{\em calculate} $P$ at 100 GeV/c without further data fitting. The model  predicts
the analyzing power at any energy, and we will examine this a little by checking on
the variation between 21.7 GeV/c and 24 GeV/c and predicting the $A_N(t)$ at 250
GeV/c. The prediction for the $I=0$ part of the $pp$ analyzing power through the
RHIC colliding beam range will be given and some model dependence examined.

We will attempt to include the $\rho$ and $a_2$ in our
analysis  in order to apply our results to  $pp$ scattering . The data from FermiLab E704
 \cite{E704} will be used to make tentative predictions for $pp$ polarimetry at high energy.

The effects we are looking at are rather small and the errors involved in the
analysis are significant and generally correlated. Therefore we devote the last
section of the paper to a careful analysis of the errors of the Reggeon spin-flip couplings 
determined earlier in the paper. The implications for the errors on the analyzing
power as a function of energy are also worked out.

\section{}
The formula we use for the $pC$ analyzing power was derived by Kopeliovich and the
author \cite{K&T}. The approximations that go into the formula are discussed in that
paper, and it is believed to be quite reliable over the small $t$-range involved in
the CNI experiments. A particularly transparent way to write it is
\begin{equation} \label{analyzingpower}
\frac{A_N(t,\tau)}{A_N(t,0)} = 1 - \frac{2}{\kappa}Re[\,\tau(s)]  +
\frac{2}{\kappa} Im[\,\tau(s)] f(t),
 \end{equation}
 with
 \begin{equation}
 f(t)=\Bigl((1 +\rho_{pC}^2(t))
(t/t_c) (F_C^h(t)/F_C^{em}(t)) -
\rho_{pC}(t)-\delta_{pC}(t)\Bigr)/(1-\rho_{pC}(t)  \delta_{pC}(t))
\end{equation}
$\tau(s)$ denotes the hadronic spin-flip parameter for the scattering. It will be defined
precisely below. $A_N(t,\tau)$ is the analyzing power in question and $A_N(t,0)$ is
``pure" CNI, the analyzing power in the absence of hadronic spin-flip. 
$F_C^{em}(t)$ is the electromagnetic form-factor and $F_C^h(t)$ is
the hadronic form-factor for carbon; these are calculated in \cite{K&T}.
$\kappa = 1.79$, $t_c=-8 \pi Z\alpha/\sigma^{pC}_{tot}$ and 
$\rho_{pC}(t)$ denotes the ratio of real to imaginary parts of the $pC$
amplitude (It depends on $t$ even if $\rho$ for $pp$ does not, as assumed in the
derivation of Eq.1; it is also calculated in
\cite{K&T}.) $\delta_{pC}(t)$ denotes the Bethe phase \cite{K&Ta} for proton-carbon
scattering; it is important only at the smallest values of $|t|$, smaller than for
which data so far exists, but we carry it along anyway. We have neglected in writing
Eq.\ref{analyzingpower} the dependence of the differential cross-section on $\tau$ which is insignificant when $\tau$ is small, in the range we anticipate.
$\tau(s)$ is defined by
\begin{equation} 
g_5(s,t) = \tau(s) \frac{\sqrt{-t}}{m} g_0(s,t) 
\end{equation}
$g_0(s,t)$ and $g_5(s,t)$ denote, respectively, the spin independent and spin-flip
$pC$ elastic amplitudes. $m$ denotes the proton mass. One of the main results of
\cite{K&T} is that $\tau(s)$ is equal to the $I=0$ part of the corresponding
spin-flip factor for $pp$ scattering. It is, in general, complex and
depends on $s$ in an unknown way, but its
$t$-dependence can be neglected over this small range of $t$. All the important
$t$-dependence of Eq.1 comes from the variation of the form factors and of
$\rho_{pC}$. The resulting dependence of $f(t)$ is shown in Fig.1.

It is important to note that this formula is given in terms of $\tau(s)$ rather than
\[r_5^{pC}(s,t)=\tau(s) (i + \rho_{pC}(s,t))\] which is sometimes used. This has the
advantage that, by the theorem of \cite{K&T}, $\tau$ is independent of $t$ while
$r_5^{pC}$ has $t$-dependence inherited from $\rho_{pC}$.
\begin{figure}[htbp]
\begin{center}
\includegraphics{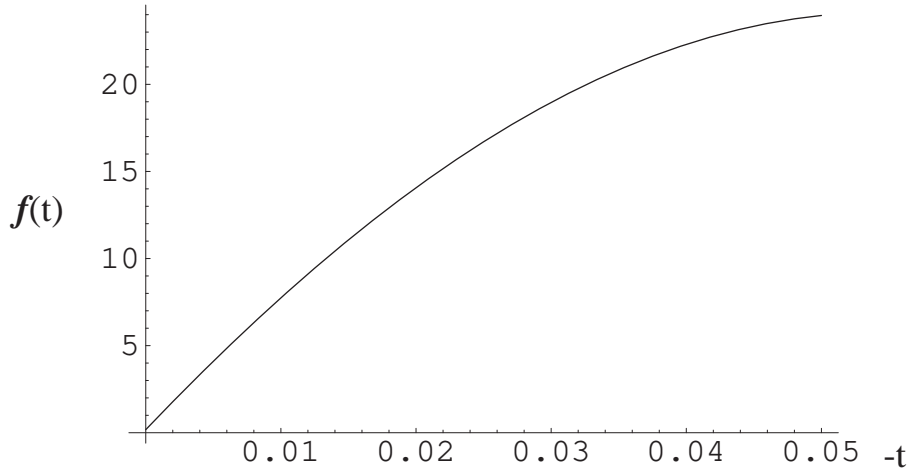}
\caption{{\it The t-dependent coefficient in the second term of Eq. 1.}}
\label{f(t)}
\end{center}
\end{figure}

The E950 group fit their data in \cite{E950} and determined values for $r_5^{pC}$.
We could simply use their results, but because the propagation of errors to our
later results is important, we have done a linear regression analysis of our own and
calculated the error matrix for $Re(\tau)$ and $Im(\tau)$. The data we fit is given
in their Table 1 and we used the errors given there to determine the weights in the
regression analysis. These errors include an approximately 12\% error in the AGS
beam polarization. (We have followed E950 in adding the errors linearly; perhaps this should be reexamined.) The results of this fits gives

\begin{equation} \label{taufit}
\tau=(-0.214 \pm 0.236) - (0.054 \pm 0.015) i
\end{equation}

The very large error on the real part of $\tau$ results from a modest error on the
first term in Eq.1 of about 25\%,. The
error in the imaginary part is also about 20\%. These errors are  too large
for the job of doing a 5\% absolute polarization measurement, and they will follow us
through all of this work. 

There are additonal errors in these numbers because of unassessed errors in the
parametrization of Eq.1. In particular, the value of $\rho_{pC}$ used is calculated.
It would be good to measure it. The same is true of the $pC$ differential cross section which determines $F_C^h(t)$. The error in our calculation of these quantities is probably
small, but it depends on, as input, the poorly known value of $\rho$ for $pp$ and $pn$
at 21.7 GeV/c. This will need to be addressed when the other errors are reduced.
(The values published in \cite{E950}  agree  within errors with  Eq.(\ref{taufit}).)

\begin{figure}[thb]
\centerline{\epsfbox{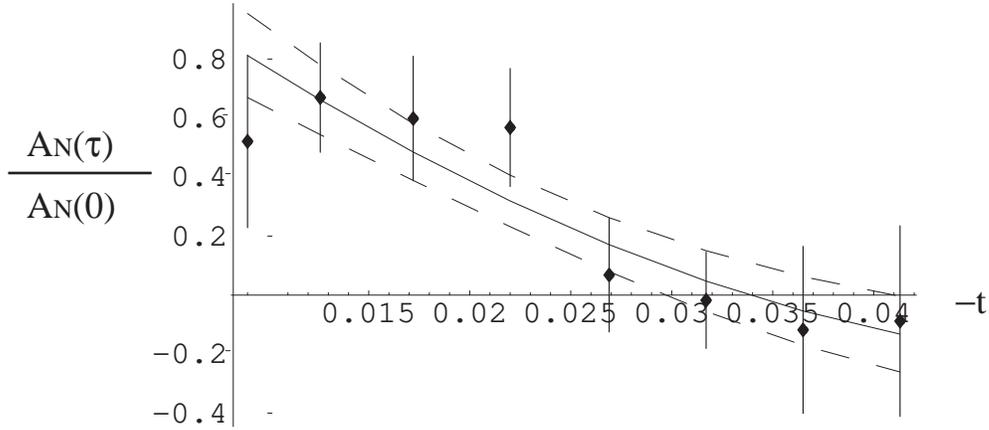}}
\medskip
{\caption[Delta]{\it The error bands for the fit of E950 data to Eq.1}
\label{error bands}}
\end{figure}
The results of the fit are shown in Fig.2 and Fig.3, first showing the error bands
on the regression analysis and second showing the $A_N(t)$ compared to the
data. 
\begin{figure}[h]
\centerline{\epsfbox{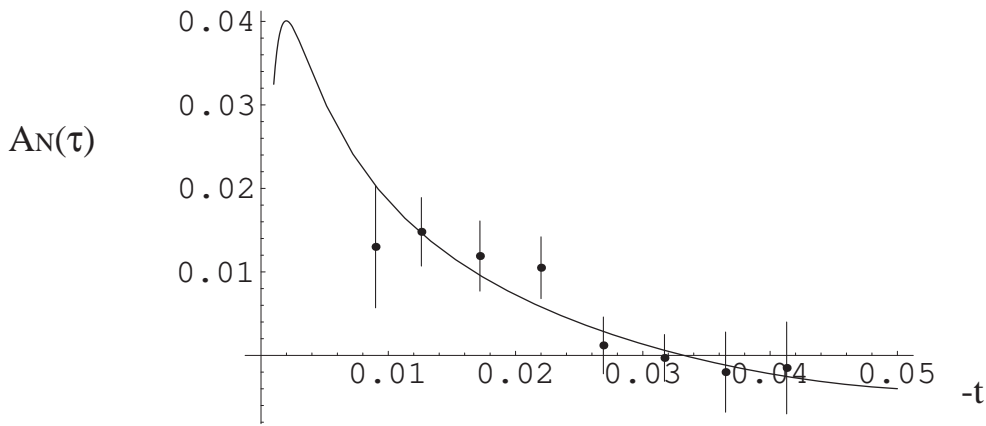}}
\medskip
{\caption[Delta]{\it Fit of analyzing power using $\tau = -0.214 -0.054 i$ to the
data of E950}
\label{950 fit}}
\end{figure}
The error matrix for  $Re(\tau)$ and $Im(\tau)$ 
is given by
\begin{equation} \label{errormatrix} 
\sigma ^2(21.7)=\left( \begin{array}{clcr}
0.0559& 0.00332\\
0.00332& 0.000213
\end{array} \right)
\end{equation}
and the corresponding 68.3\% confidence level ellipse for the fit parameters ({\em not} the $\tau$'s) is shown in Fig.\ref{950 ellipse}
\begin{figure}[thb]
\centerline{\epsfbox{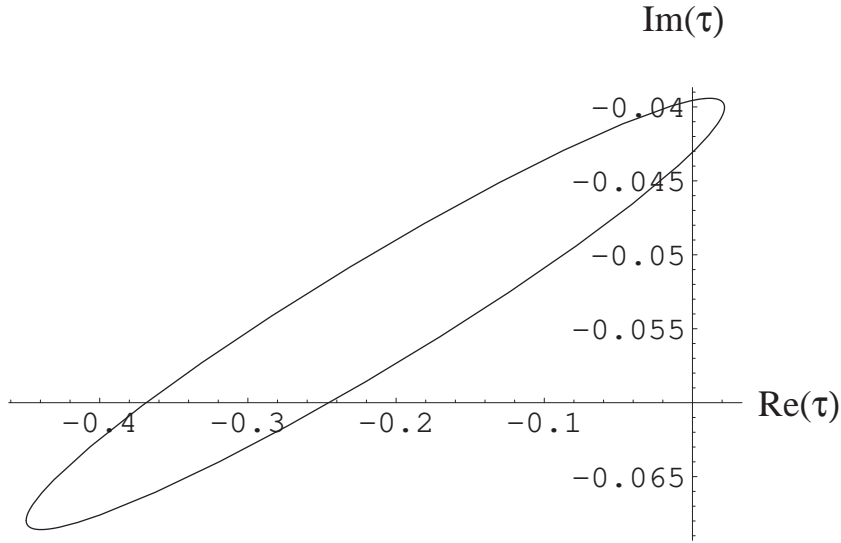}}
\medskip
{\caption[Delta]{\it The $1\sigma$ error ellipse for the fit to E950 data}
\label{950 ellipse}}
\end{figure}

It is important to bear in mind that if one attempts to fit the data using an incorrect assumption regarding the polarization, call it $P'$, one will obtain a perfectly good fit with exactly the same $\chi^2$ as the correct one; one will simply determine values of the fit parameters which scale as
\begin{eqnarray}
1 - \frac{2}{\kappa} Re(\tau')&=& \frac{P}{P'} \left (1-\frac{2}{\kappa }Re(\tau) \right) ,\nonumber \\
Im(\tau')&=& \frac{P}{P'}Im(\tau) .
\end{eqnarray}

\section{}
As a first application of these results, let us use the analyzing power just
determined at 21.7 GeV/c to determine the polarization of the beam in RHIC after
injection at 24 GeV/c. To do this we assume that $A_N(t)$ doesn't change
significantly over this small energy range. We  will return to this question after we
develop a model for the energy dependence. We will use {\em the preliminary data for the raw asymmetry} $\epsilon$$(t)$ which does not depend on any knowledge of the polarization of  the  beam, as presented
at Prague in July 2002 \cite{Prague} and at Spin 2002 \cite{Spin 2002}. The 
data for
  $\epsilon$$(t)$ with errors $e(t)$ are given in Table 1. The errors given assume that the systematic errors are equal to the statistical errors and they are combined in quadrature. (I thank D. Svirida for this information.) Evidently, these errors do not contain a
contribution from the polarization because this is the raw asymmetry, not
$A_N$.  (The values of $A_N$ given in those references are calculated assuming $P=0.27$ and so if one fits $A_N$ instead of  $\epsilon$$(t)$  one will obtain values for $\tau$ scaled as mentioned at the end of the last section. These turn out to be very different than the values in the E950 fit.)

 At this moment, the data is not yet published and must be considered in the
context of conference presentations. In the not-too-distant future, new data will become available and this analysis may need to be revised. We choose to present this work based on the preliminary results  for two reasons: (1) In order to illustrate the method we propose, it is necessary to apply it to real data, and (2) the results should be immediately useful to the ongoing program at RHIC in order to better understand the polarization at different energies.
\begin{table}[h]
\centering
$\begin{array}{|c|c|c|}
\hline
\mbox{-$t$} & \mbox{$\epsilon$$(t)$}& \mbox{$e(t$)} \\ 
\hline
0.0117 & 0.0064 & 0.00060 \\
0.0138 & 0.0059 & 0.00043 \\
0.015 &
0.0046 & 0.00068\\ 0.0184 & 0.0019 & 0.00076
\\ 0.0194 & 0.0024 & 0.00054 \\
0.0217 & 0.0025 & 0.00087 \\
0.0250 & 0.0016 & 0.00068 \\ 
0.0251 & 0.0033 & 0.00098 \\
0.0306 & 0.00005 & 0.00081 \\
0.0361 & 0.00035 & 0.00098 \\
0.0417 & -0.00046 & 0.00116 \\
0.0474 & 0.00089 & 0.00138
\\ \hline 
\end{array}$
\caption{\sl -t, raw asymmetry $\epsilon$$(t)$, and errors $e(t)$ for RHIC 24
GeV/c}
\end{table}
 
\begin{figure}[t]
\centerline{\epsfbox{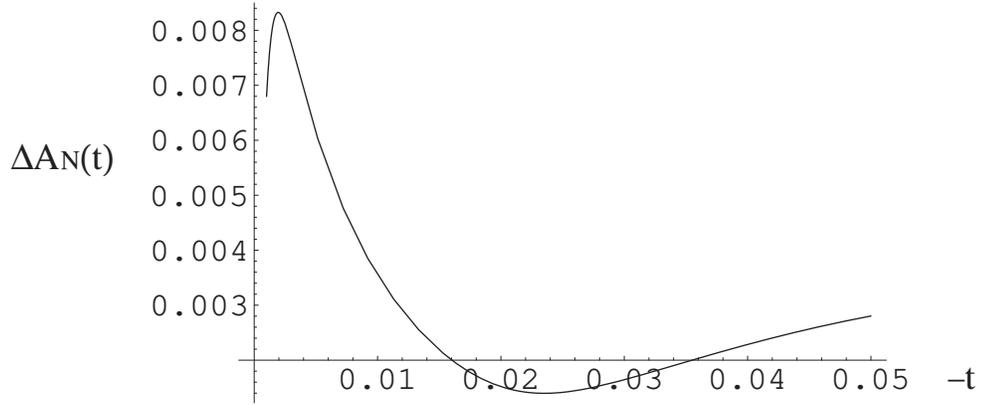}}
\medskip
{\caption[Delta]{\it The uncertainty in the analyzing power predicted at 24 GeV/c}
\label{deltaAn}}
\end{figure}

Starting from the relation
\[{\epsilon}(t) = P A_N(t), \] we minimize the chi-square
 \begin{equation} \label{chisq(24)}
{\chi^2(24)}= \sum_{i=1}^{i=12}(\epsilon(t_i)- P A_N(21.7,t_i))^2/e(t_i)^2.
\end{equation}
$A_N(21.7,t_i)$ denotes the form in
Eq.(\ref{analyzingpower}) with
$\tau=-.0214 - .054 i$ evaluated at $t_i$. This is easily done with the result
that
\begin{equation}
 P(24)= 0.39 \pm 0.02.
\end{equation}
The error here is just the statistical error of the fit. There is a larger error
associated with the error in the determination of $A_N(t)$. It can be determined using
the formula for $A_N(t)$ and the error matrix, Eq.(\ref{errormatrix}) to be about $\pm
0.068$ or about 17\%. See Fig.\ref{deltaAn}
 \begin{figure}[b!]
\centerline{\epsfbox{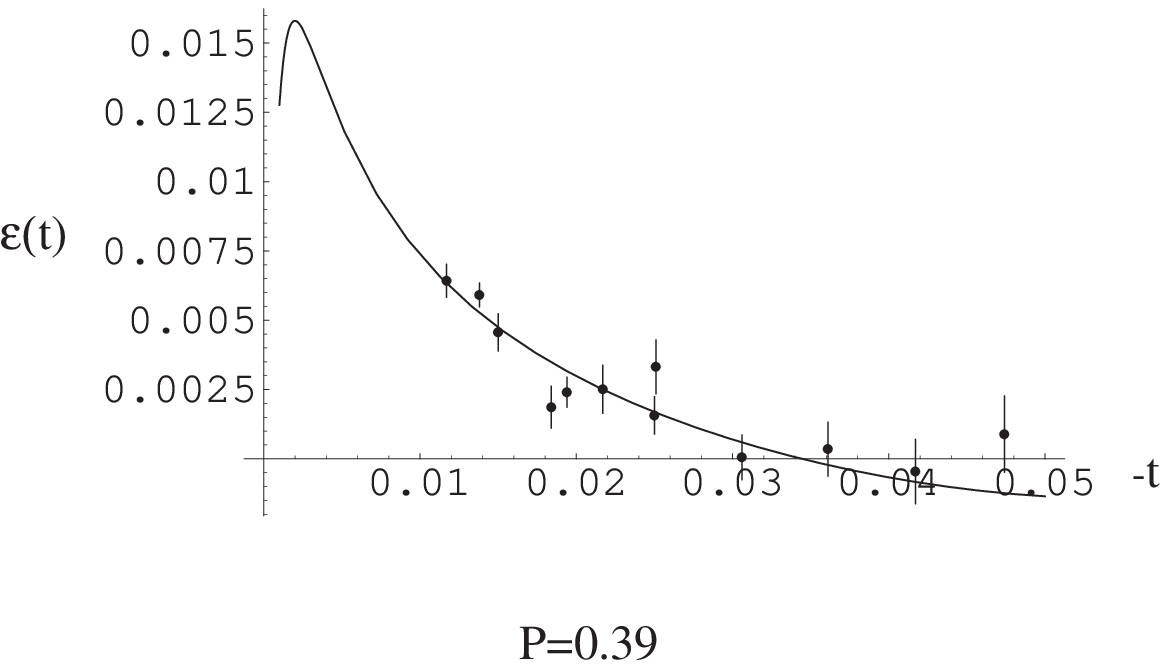}}
\medskip
{\caption[Delta]{\it Raw asymmetry at 24 GeV/c  as measured and as predicted by E950 analyzing power
with P=0.39}
\label{asym24}}
\end{figure}

Now using the E950 value of $\tau$ and this polarization we plot together the
expected asymmetry along with the RHIC 24 GeV/c data, Fig.\ref{asym24}. The fit is
reasonable with  $\chi^2$ = 1.5 per dof. 
\newpage
\section{}
Can we say anything about the energy dependence of $A_N(t)$ based on the the E950 or
RHIC 24 data alone? The simplest plausible Regge model to describe the energy
dependence has three Regge poles: the Pomeron, the $f_2$ and the $\omega$. There have
been at least two groups \cite{Cudell, Block} which have produced fits to the extensive
data for
$pp$ and
$\bar{p}p$ elastic scattering over an enormous energy range which includes almost all
of the RHIC energy range. The lowest energy, which includes our starting point, is
marginal to the fits. Nevertheless we will use these models as a basis for our own
model. In these fits the $f_2$ and the $\omega$ are always treated as simple Regge
poles, but the more complex nature of the Pomeron is taken into account by allowing
it to be alternatively a simple pole a little above $J=1$ , or multiple poles at $J=1 $ in
order to produce single or double logarithms in the asymptotic energy dependence of the
total cross section. For this work we have chosen two representitive fits from
\cite{Cudell}, a simple pole at 1.0933 or a multiple pole at 1 designed to give a
log-square growth. We will work through the first case first and then indicate how
the second differs from it.

We begin with the parametrization of $pp$ elastic scattering given
by Cudell et al \cite{Cudell}, though one might do the same thing using other
parametrizations such as that of Block et al \cite{Block}. Since it is
known that the elastic, non-flip scattering is overwhelmingly $I=0$
exchange, even at 24 GeV/c \cite{np data}, we will assume the Regge
couplings that they determine are for the $I=0$ families and so directly
applicable to $pC$ scattering. The form they assume for the forward
amplitude is
\begin{equation}
g_0(s,0)= g_P(s) + g_f(s) + g_\omega (s)
\end{equation}
with
\begin{eqnarray} \label{Cudellmodel}
g_P(s)&=&-X s^\epsilon (\cot\!\frac{\pi}{2}(1 + \epsilon) - i),\\ \nonumber
\newline g_f(s)&=&-Y s^{-\eta} (\cot\!\frac{\pi}{2}(1 - \eta) - i), \\ \nonumber
\newline g_\omega(s)&=&- Y's^{-\eta'} (\tan\!\frac{\pi}{2}(1 - \eta') + i) 
\end{eqnarray}
normalized that $Im(g_0(s))= \sigma_{\rm tot}(s)$. The values of the
parameters given by them are

\begin{eqnarray}
 \epsilon= 0.0933,\, \eta =0.357, \,\eta' =0.560, \\ \newline
 X=18.79,\, Y=63.0,\, Y'=36.2,\nonumber
\end{eqnarray} with $X, Y, Y'$ in $mb$.

Our model assumes that the spin-flip $pp$ $I=0$ exchange
amplitude $g_5(s,t)$ is given by

\begin{eqnarray} 
g_5(s,t) &=& \tau(s) \frac{\sqrt{-t}}{m} g_0(s,t) \\ \nonumber
{} &=& \frac{\sqrt{-t}}{m} \{\tau_P\, g_P(s) + \tau_f\, g_f(s) +
\tau_\omega\, g_\omega(s)\}.
\end{eqnarray}
where $\tau(s)$ depends on energy but not on $t$ over the CNI range. It is in
general neither real nor constant in $s$ and is given by

\begin{equation} \label{tau(s)}
\tau(s) = \{\tau_P\, g_P(s) + \tau_f\, g_f(s) + \tau_\omega\,
g_\omega(s)\}/g_0(s,0)
\end{equation}
 where the $\tau_i$'s are energy-independent, {\em real} constants. The
phases of the amplitudes come only from the energy dependence as given in
Eq.(\ref{Cudellmodel}). This is the key assumption from Regge theory which we need: as
a result the real and imaginary parts of $\tau(s)$ are given at each energy
in terms of the three real constants $\tau_P, \tau_f$ and $\tau_\omega$.

\begin{figure}[thb]
\centerline{\epsfbox{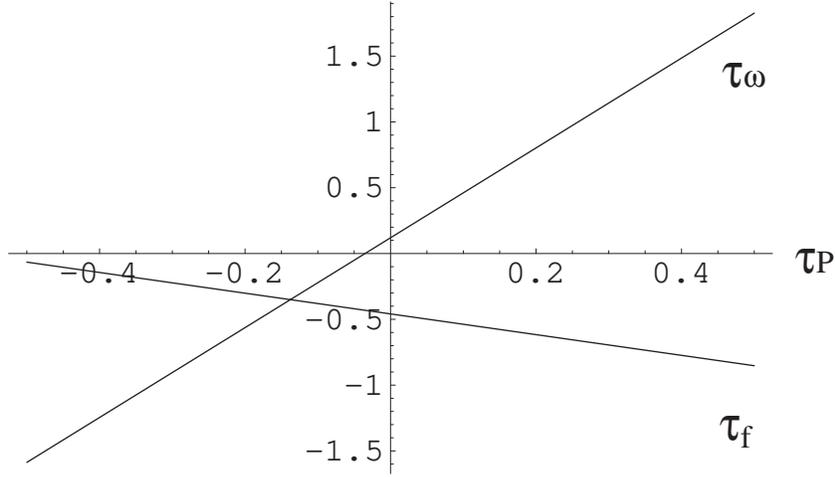}}
{\caption[Delta]{\it Relations between the three Regge spinflip coupling based on E950 results alone}
\label{intersection}}
\end{figure}

Since from E950 we can determine only two real parameters, it is plain that we can
not fix this model completely. The best we can do is obtain a relation between the
three spin flip couplings; for example, in Fig.\ref{intersection} we plot
$\tau_f$ and $\tau_{\omega}$ as constrained by the values of Eq.(\ref{taufit}).

The recent runs at RHIC have also provided data at 100 GeV/c. The polarization is not
known there, so is it possible to obtain useful information from it? The answer is
yes: if the asymmetry is described by the CNI formula, which is our
fundamental assumption, then a fit to the raw asymmetry will determine $ (1
- \frac{2}{\kappa }Re[\tau(100)])P(100)$ and $\frac{2}{\kappa}Im[\tau(100)] P(100) $.  (In this and the following sections we will write the argument of the shape $S$ and $\tau$ as the lab momentum rather than the corresponding $s$. When we briefly discuss colliding beams, we will revert to $s$, but there should not be any confusion.) Thus
the ``shape" $S$ of the distribution
\begin{equation}
S(p_L)= \frac{Im[\tau(p_L)]}{\kappa/2 - Re[\tau(p_l)]}
\end{equation}
can be determined at any energy without knowing $P$. Now once we have the three
quantities,
$Re[\tau(21.7)], Im[\tau(21.7)]$ and $S(100)$ we have three linear equations in
$\tau_P, \tau_F$ and $\tau_{\omega}$ and can determine them all. This then will give
us $\tau(p_L)$ at $p_L=100$ GeV/c  and thereby we can calculate the polarization
$P(100)$. We will now go through this process.

Using data from the same sources as used at 24 GeV/c \cite{Prague, Spin 2002}, %
\begin{table}[h]
\centering
$\begin{array}{|c|c|c|}
\hline
\mbox{-$t$} & \mbox{$\epsilon(t)$}& \mbox{$e(t$)} \\ 
\hline
0.0117 & 0.0036 & 0.00055 \\
0.0138 & 0.0029 & 0.00047 \\
0.015 & 0.0034 & 0.00060 \\ 
0.0184 & 0.0018& 0.00069 \\ 
0.0194 & 0.0014 & 0.00058 \\
0.0217 & 0.0025  & 0.00087 \\
0.0249 & -0.0004 & 0.00072 \\ 
0.0251 & 0.0009 & 0.00085 \\
0.0306 & -0.0010 & 0.00072 \\
0.0360 & 0.0010 & 0.00091 \\
0.0416 & 0.0013 & 0.00116 \\
0.0473 & -0.003 & 0.00146
\\ \hline 
\end{array}$
\caption{\sl -t, raw asymmetry $\epsilon$$(t)$, and errors $e(t)$ for RHIC 100
GeV/c}
\end{table}

We then fit this as in Section 2 to the formula in Eq.(\ref{analyzingpower}) with the
right hand side multiplied by the unknown $P(100)$. There is a small calculable
energy dependence to $f(t)$, and it is taken into
account in the fit. The result of the regression is 
 \begin{eqnarray}P(100)
(1- \frac{2}{\kappa} Re[\tau(100)])&= &0.263  \\ \nonumber 
 P(100) \frac{2}{\kappa }Im[\tau(100)]&=& -0.0137.
 \end{eqnarray}
 Combining these together we get for the shape of
the distribution
\begin{equation} \label{shape100}
S(100)= -\frac{0.0137}{0.263}=-0.052
\end{equation}
By using Eq.(\ref{taufit}) the value of $\tau(21.7)$ as determined in E950 in terms of
the Regge spin-flip couplings via Eq.(\ref{tau(s)}) and in the same way express
$S(100)$ in term of the Regge spin-flip couplings via $\tau(100)$ and
Eq.(\ref{tau(s)}), we have three equations to solve with the result
\begin{eqnarray} \label{reggecouplings}
\tau_P &=& -0.02 \\ \nonumber
\tau_f &=& -0.43 \\ \nonumber
\tau_{\omega} &=& 0.03
\end{eqnarray}
There are significant errors in these determinations and we will return to them in
the next section.
 
\begin{figure}[thb]
\centerline{\epsfbox{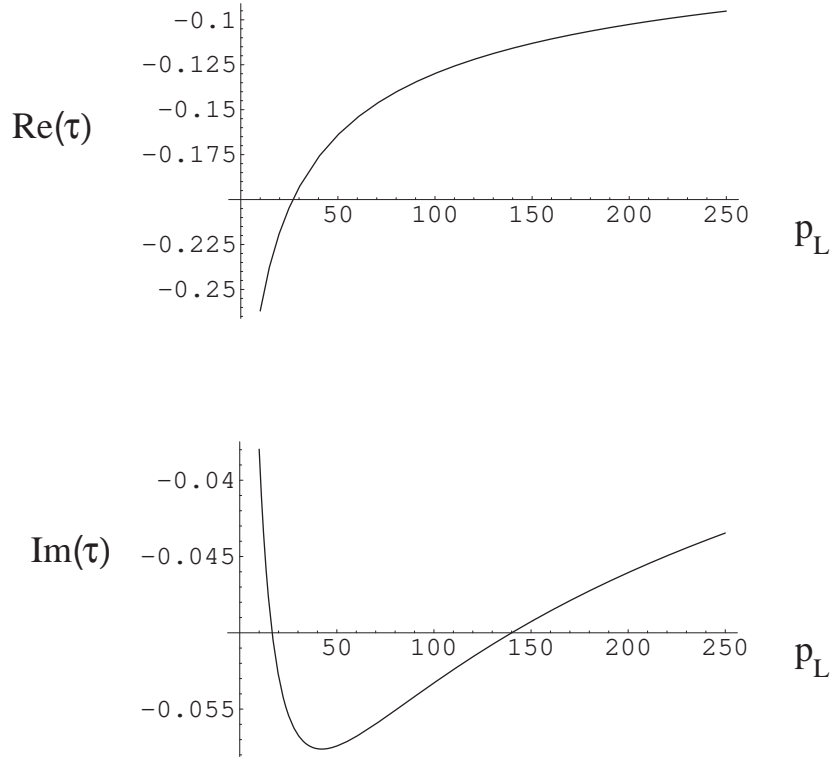}}
\medskip
{\caption[Delta]{\it Predicted s-dependence of $\tau $}
\label{energydep}}
\end{figure}

 These results allow us to calculate $\tau(s)$ at any higher energy. (The model as it stands, is not really suitable for going to lower energy because lower lying Regge poles will rapidly beome important. Thanks to Boris Kopeliovich for emphasizing this limitation \cite{Berger}, \cite{Kramer}.) Fig. \ref{energydep} shows the $p_L$ dependence of the real and imaginary parts of $\tau$ over the RHIC fixed target range. While the spin-flip couplings are evidently becoming smaller with increasing  energy, there  remains significant hadronic spin-flip at the top energy.

\begin{figure}[thb]
\centerline{\epsfbox{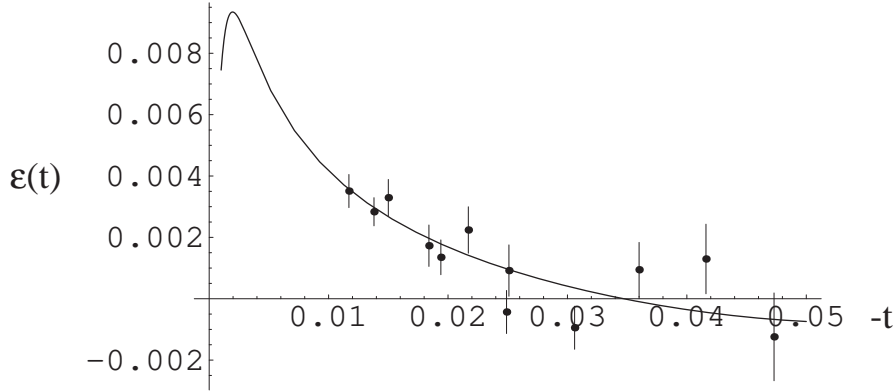}}
\medskip
{\caption[Delta]{\it Raw asymmetry at 100 GeV/c predicted by model $\tau(100)$
and polarization predicted to be 0.23}
\label{asym100}}
\end{figure}

We will use these results first for  $\tau(100)$ and thereby determine
the polarization at 100 GeV/c, $P(100)$. 
From Eq.(\ref{tau(s)}) we find
\begin{equation} 
\tau (100) = -0.130 - 0.053 i 
 \label{tau(100)}
\end{equation}
With  $\kappa/2 - Re[\tau(100)]$ and/or $Im[\tau(100)]$, we can use the measured values of  $P(100) (\kappa/2 - Re[\tau(100)])$ and/or $P(100) Im[\tau(100)]$ to quickly determine $P(100) =0.23$.  Alternatively, as in Section 3, we can fit the 12 measured asymmetry values to $A_N$ using Eq.(\ref{tau(100)}). Then minimizing the $\chi ^2$ as in Eq.(\ref{chisq(24)}) we find
\begin{equation}
P(100) = 0.23 \pm 0.02,
\end{equation}
again the error is just the statistical error of this fit.  In Fig.\ref{asym100} we show the raw asymmetry measured at 100 GeV/c plotted with the prediction using  Eq.(\ref{tau(100)})  and this value of $P$.
The agreement is reasonable,  with $\chi^2$ about 1.5/dof, about the same as the fit at 24 GeV/c. This is rather nice support for our approach.

Given that $P(24)$ was found to be 0.39, the value found for $P(100)$ is surprisingly small. There seem to be
three possiblilities to explain this: (1) There could be significant depolarization in
the RHIC acceleration to 100 GeV/c. This is not expected, but this may be a signal of a problem in the acceleration. 
(2) The model used for energy dependence could be
wrong. This seems very likely in its details, but if one compares the raw asymmetries
at 24 and 100 GeV/c, one sees that a very large drop in the analyzing power--nearly a
factor of 2 from 24 to 100 GeV/c is required if $P$ remained constant. Of course,
there might be a problem with the data at the smallest $|t|$,  but it is not
suggested by the errors assigned.

In Section 3 we determined $P(24)$ by assuming that $\tau (24)$ is the same as $\tau
(21.7)$ determined by E950. Now we have a prediction for $\tau(24)$ and it is a
little different:
\begin{equation}
\tau(24)= -0.207 - 0.055 i.
\end{equation}
We have checked its significance in determining $P$ via Eq.(\ref{chisq(24)}) and find the
best fit at $P=0.41 \pm 0.02$, slightly different but certainly well within errors.
(Remember that the result of the fits at 24 GeV/c were not used in the energy
dependence calculations, so this is another check, although not a very strong one.)

\begin{figure}[bht]
\centerline{\epsfbox{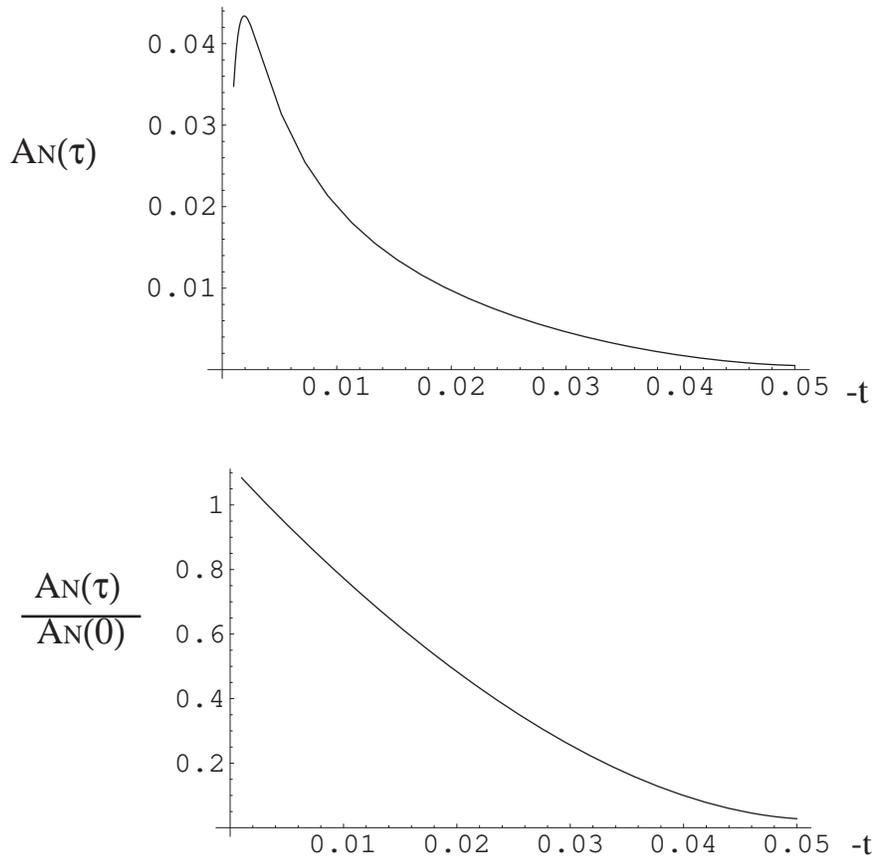}}
\medskip
{\caption[Delta]{\it  Analyzing power predicted for 250 GeV/c  pC scattering, and its ratio to the pure CNI analyzing power at 250 GeV/c}}
\label{An(250)}
\end{figure}

In the near future it is hoped to have a polarized proton run at RHIC at $p_L= 250$ GeV/c. The predicted analyzing power is shown in  Fig.10.  It is a little larger and has a slightly different shape from the 21.7 GeV determination. For clarity we also show
in that figure the ratio of predicted to pure CNI analyzing power.  This shows that the hadronic spin-flip must  be taken into account in $A_N$ even at the highest RHIC beam energy on a fixed target.

\begin{figure}[!thb]
\centerline{\epsfbox{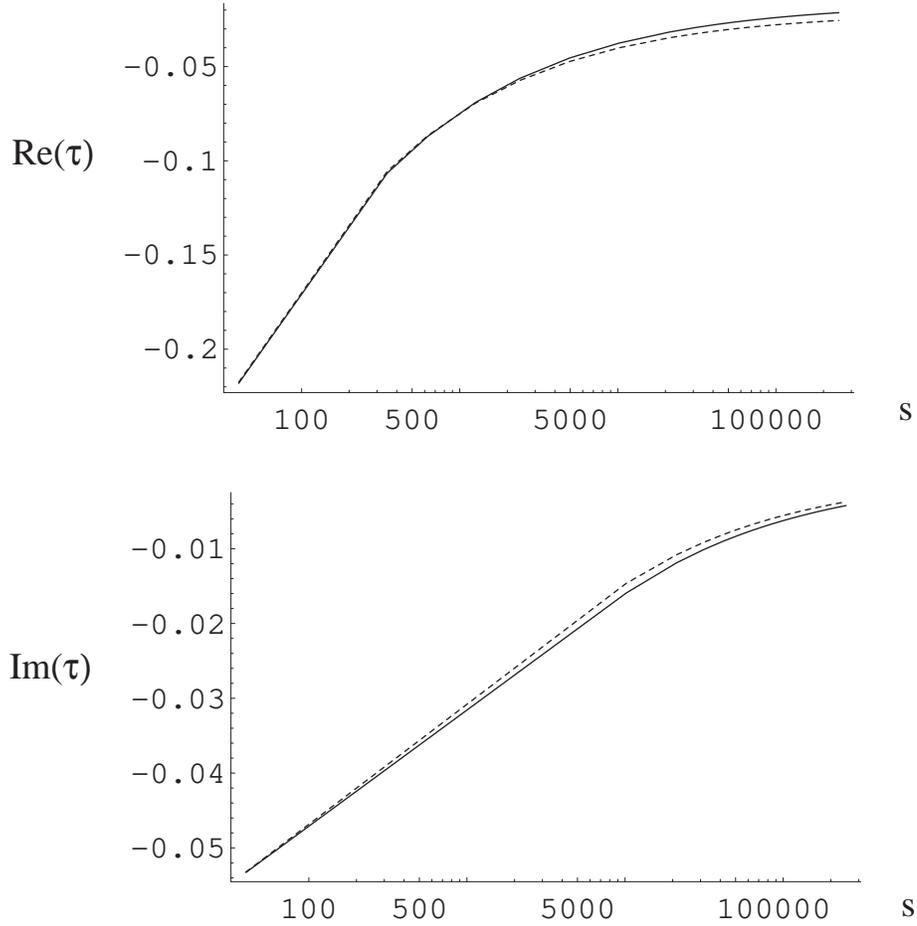}}
\medskip
{\caption[Delta]{\it The I=0 part of the pp $\tau(s)$ plotted here against s
through the RHIC colliding beam energy range. The solid line is the prediction of the 3-pole model, the dashed line the model with log-squared growth.}
\label{full RHIC range}}
\end{figure}

As mentioned earlier, we also examined the behavior of a Cudell et al \cite{Cudell}
fit with a log-squared asymptotic behaviour for the Pomeron. The fit we chose to look
at has the Pomeron form replaced by
\begin{equation}
g_P(s)= i(A + B \log^2{s}) + \pi B \log{s},
\end{equation} $s$ in units of GeV$^2$.
The Regge pole forms are unchanged, but the associated parameters are somewhat
different:
\begin{eqnarray}
A=25.29, B=0.227,  \eta =0.341, \,\eta' =0.558 ,\\ \newline
Y=52.6,\, Y'=36.0\nonumber,
\end{eqnarray} with $A, B, Y, Y'$ in $mb$. The spin-flip parameters associated with
this form for the unpolarized amplitudes are found to be, in the same way as for
the 3-pole model
\begin{eqnarray}
\tau_P'&=&-0.02 \nonumber \\
\tau_F'&=&-0.49 \nonumber \\
\tau_{\omega }' &=& 0.02.
\end{eqnarray}
These numbers are very similar to those found for the three pole model but not
really directly comparable because the models are, in principle, quite different.
They lead to very little difference in the model preditions. We can accentuate the
difference by looking at the very high energy behavior where the difference
between the power and the log's should be greatest, so in
Fig.\ref{full RHIC range} we show the behaviour of $Re[\tau]$ and $Im[\tau]$
over the full RHIC colliding beam energy range. (Note that here the $x$-axis is the
$s$ axis.) The difference remains tiny over the entire range. In either case the $I=0$ hadronic spin-flip should be negligible for $5 \%$ polarimetry for colliding beams above about 70 GeV/c.

\section{}
The last point leads us to make a brief observation. The $\tau$ shown in
Fig.\ref{full RHIC range} is only the $I=0$ part of the spin-flip, as predicted
from the proton-carbon analysis. In order to use this for high-energy $pp$
scattering we must add in the $I=1$ spin-flip, in Regge terms the $\rho$ and the
$a_2$, which are expected to be quite large \cite{Berger}.  If we fit the E704
data \cite{E704, Buttimore} we find \[\tau(E704) = 0.2 \pm 0.2 + 0.024 i \pm 0.020 i\] to be
compared with our prediction of the $I=0$ part at $p_L=200 \, GeV/c$
\[ \tau_0(200) = -0.10  - 0.046 i .\]  Even with the very large error on the E704 result, these numbers are incompatible. The difference can be naturally explained by
adding in large $I=1$ spin-flip contributions which are absent in the $pC$ scattering but are known to be important at lower energy in $pp$ scattering \cite{Berger, Kramer}. If we assume that the
$\rho$ and $a_2$ have the same Regge behaviour as the $\omega$ and the $f_2$,
respectively, we can determine the $C=-1$ and the $C=+1$ combined $I=0$ and $I=1$
Regge flips to be,
\[ \tau(C=-) = -2.10 \] and  \[ \tau(C=+) = 0.69 .\] The relation of these numbers to the $\rho$ and $a_2$ couplings requires more study. However, just with these numbers and the value we have determined for $\tau_P$ we can calculate the values of the proton spin-flip expected at high RHIC colliding beam energy, and we find $\tau_{pp}(70^2)= 0.04 + 0.025 i$ and $\tau_{pp}(500^2)=-.01 + 0.006 i$ so the hadronic spin-flip will have little effect on $A_N$ in the CNI region at the highest energies.
\section{}
\begin{figure}[!thb]
\centerline{\epsfbox{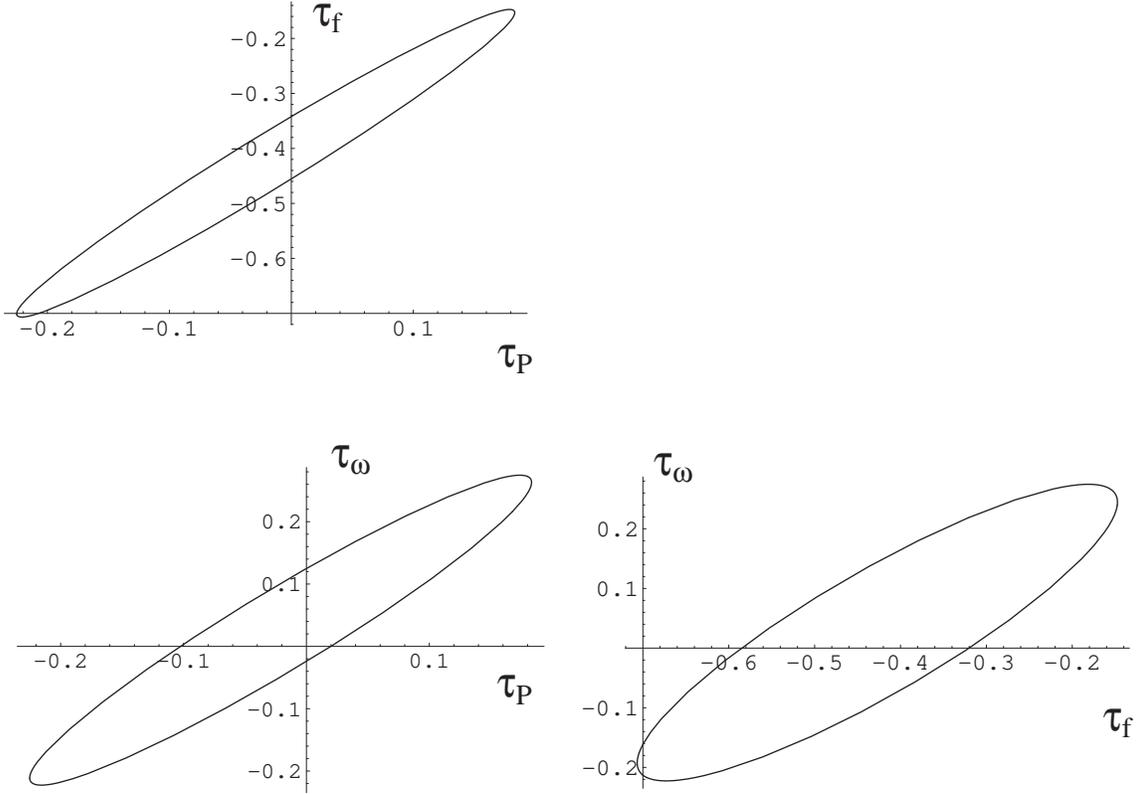}}
\medskip
{\caption[Delta]{\it Three projections of the error ellipsoid for the three I=0
Regge couplings}
\label{ellipses}}
\end{figure}

Finally, we would like to make some comments on how well the Regge parameters are
determined in this model and using this data,  especially on the question of
the Pomeron spin-flip coupling. Starting from the error matrix for $\tau(21.7)$
and error on $S(100)$, which requires the error matrix for $\tau(100)$ we can easly
propagate the errors to an error matrix of the $I=0$ Regge spin-flip factors. For
completeness we give the error matrix for $\tau(100)$ determined by our fit, and
assuming the polarization $P(100)=0.23$:
\begin{equation} \label{errormatrix100}
 \sigma ^2(100)=\left( \begin{array}{clcr}
0.0164 & 0.0012\\
0.0012 & 0.0001
\end{array} \right).
\end{equation}
This leads to an error in the observed shape  $S(100) = -0.052 \pm .004$. Using
this plus Eq.(\ref{errormatrix}) leads to the corresponding $1\sigma$ error matrix for the
Reggeons
\begin{equation} 
 \left( \begin{array}{clcr}
 &\; \; \; \tau_P & \; \; \tau_f &\tau_{\omega} \; \; \;       \\ \nonumber
\tau_P & 0.0416 & 0.0559 & 0.0484 \\ \nonumber
\tau_f & 0.0559 & 0.0782 & 0.0612 \\ \nonumber
\tau_{\omega} & 0.0484 & 0.0612 & 0.0618 \\
\end{array} \right)
\end{equation}
This may be visualized with the aid of three ellipses of projections of this
ellipsoid onto the three planes defined by $\tau_P,\tau_f$ and $\tau_{\omega}$, in
Fig. \ref{ellipses}.

The errors are clearly very strongly correlated.  Using these figures,
one can respond to the question long asked
\cite{Buttimore}, is the Pomeron spin-flip coupling zero? Examination of these
figures (or the equation for the complete 3-d ellipsoid) shows that the point 
\begin{eqnarray}
\tau_P &=& 0 \\ \nonumber
\tau_f &=& -0.4 4\\ \nonumber
\tau_{\omega} &=& 0.09
\end{eqnarray}
is comfortably with all the ellipses.
So the Pomeron can have a zero spin-flip coupling, but the $f$ cannot. Somewhat
surprisingly, even at the highest RHIC energy and even in the absence of a Pomeron
spin-flip, the $pp$ spin-flip should not be expected to vanish although it is
expected to be down at the few percent level. 

We can use the error matrix for the Regge couplings to calculate the errors in the predicted $\tau(s)$ throughout the RHIC range. The errors vary slightly with energy but remain very close to 20\% of either $ (\kappa/2 -Re[\tau(s)])$ or $Im[\tau(s)]$ between $p_L=20\, GeV/c$ and $p_L=250\,GeV/c$. In particular, the error matrix implied  by our model at $p_L=100\,GeV/c$ is
\begin{equation} 
\sigma_{model}(100)^2= 
\left( \begin{array}{clcr}
0.0492 & 0.0025\\
0.0025 & 0.00014 
\end{array} \right),
\end{equation}
 larger than the errors of the fit Eq.(\ref{errormatrix}). At $p_L=250 \,GeV/c$ it becomes
 \begin{equation}
 \sigma_{model}(250)^2= 
\left( \begin{array}{clcr}
0.0467 & 0.0019\\
0.0019 & 0.00009
\end{array} \right).
\end{equation}

We close by emphasizing that the numerical results obtained here are dependent on the preliminary results announced at Prague and Spin 2002 and they may change significantly.   It will be very interesting to see how these results hold up when new, more precise
experiments are carried out, both at the low end of the energy range here
considered, and at the high end. It will also be important to extend the work of
Section 5 to reach stronger conclusions regarding the $I=1$ couplings and to extend the Regge model downward in energy. The present
level of accuracy on the experiments limits the strength of the conclusions we can
draw, but it does appear that the CNI pC polarimeter should be a very precise
relative polarimeter, with a modest absolute accuracy.

I would like to thank Boris Kopeliovich, Gerry Bunce, Dima Svirida and Nigel
Buttimore for very useful discussions of this physics.

\end{document}